\documentclass[a4paper,conference]{IEEEtran}

\usepackage{cite}
\usepackage{graphicx}
\usepackage{subfigure}
\usepackage{url}
\usepackage{amsmath}
\usepackage{soul}
\usepackage{array}

\usepackage{algpseudocode}
\usepackage{algorithm}
\usepackage{algorithmicx}
\usepackage[dvipsnames]{xcolor}
\usepackage{mathrsfs}

\usepackage{hyperref}
\hypersetup{
    colorlinks=true,
    linkcolor=blue,
    filecolor=magenta,
    urlcolor=cyan,
}

\graphicspath{{plots/}}

\newcommand{\ji}{\varphi}
\newcommand{\GN}{G}

\newcommand{\ZA}{{\bf Z}_A}
\newcommand{\ZB}{{\bf Z}_B}
\newcommand{\Xa}{{\bf X}_a}
\newcommand{\Yb}{{\bf Y}_b}
\newcommand{\xa}{{\bf x}_a}
\newcommand{\yb}{{\bf y}_b}
\newcommand{\Rb}{{\bf R}}
\newcommand{\Fb}{{\bf F}}
\newcommand{\nb}{{\bf n}}
\newcommand{\ab}{{\bf a}}
\newcommand{\db}{{\boldsymbol{\delta}}}
\newcommand{\xib}{{\boldsymbol{\xi}_a}}
\newcommand{\etab}{{\boldsymbol{\eta}_b}}

\newcommand{\apjl}{ApJL}

\usepackage{fancyheadings}
\begin{document}

\title{Conservation of Angular Momentum in the Fast Multipole Method}

\author{\IEEEauthorblockN{Oleg Korobkin, Hyun Lim, Irina Sagert, Julien Loiseau, Christopher Mauney, M.~Alexander~R. Kaltenborn,\\
Bing-Jyun Tsao, Wesley~P. Even}
\IEEEauthorblockA{Los Alamos National Laboratory\\
Los Alamos, NM 87544 USA}
}

\maketitle

\begin{abstract}
Smoothed particle hydrodynamics (SPH) is positioned as having ideal conservation properties.
When properly implemented, conservation of total mass, energy, and both linear and angular momentum is guaranteed exactly, up to machine precision.
This is particularly important for some applications in computational astrophysics, such as binary dynamics, mergers, and accretion of compact objects (neutron stars, black holes, and white dwarfs).
However, in astrophysical applications that require the inclusion of gravity, calculating pairwise particle interactions becomes prohibitively expensive.
In the Fast Multipole Method (FMM), they are, therefore, replaced with symmetric interactions between distant clusters of particles (contained in the tree nodes) \cite{dehnen00}.
Although such an algorithm is linear momentum-conserving,
it introduces spurious torques that violate conservation of angular momentum.

We present a modification of FMM that is free of spurious torques and conserves angular momentum explicitly.
The new method has practically no computational overhead compared to the standard FMM.
\end{abstract}

\section{Introduction}
\label{sec:intro}

To correctly model astrophysical phenomena like the orbital motion of planets, accretion disks, rotating neutron stars and black holes, compact star mergers, and galactic disks, numerical approaches have to accurately determine gravitational interactions and ensure the conservation of energy, linear, and angular momentum.
Many self-gravitating astrophysical systems are well-described by Newtonian gravity, and there is a variety of numerical methods to model the corresponding gravitational field.
For example, grid-based methods often determine the gravitational potential by solving the Poisson equation and using iterative solvers and/or Fourier transforms \cite{Fryxell2000, DSouza2006, Stone2008, Almgren2010}.
Particle-based methods, like N-body and Smoothed Particle Hydrodynamics (SPH) \cite{Hernquist1989,Springel2005,Vanaverbeke2009,Yokota2012,Loren2010,Price2018,Rosswog2020} usually rely on tree codes \cite{Barnes1986}, particle-mesh methods \cite{Bagla2002}, or the Fast Multipole Method (FMM) \cite{Greengard1997,Cheng1999,Warren1995}.
However, when it comes to angular momentum conservation, some methods, including FMM, struggle.
The latter conserves angular momentum exactly when used in zeroth-order or with a small error for low values of the so-called multipole acceptance criterion (MAC) angle. However, this comes with a high computational cost.
To reduce this performance cost, a larger MAC angle is typically chosen, which then correspondingly increases the error in angular momentum.

In this work, we investigate a propositions within the standard FMM to conserve the angular momentum \emph{by construction}, on par with linear momentum.
The spurious torques in standard FMM arise due to the pairwise action-reaction forces being slightly offset from the line connecting two particles.
Here we propose a new method, in which the misaligned pairwise forces are reprojected onto the line connecting two particles, such that spurious torques disappear.

The rest of the paper is organized as follows: in Section~\ref{sec:problem}, we describe FMM~\cite{dehnen00} and the specifics of the problem conserving angular momentum for high-order FMM.
We start by reviewing the symmetric multipole interactions as introduced in~\cite{Warren1995, dehnen00}.
In Section~\ref{sec:solution}, we describe two proposed approach, which ensures angular momentum conservation. In Section~\ref{sec:results}, we present the numerical tests, highlighting conservation of linear and angular momentum;
we conclude with the Section~\ref{sec:conclusion}.

\section{Method}
\label{sec:problem}
In this section, we briefly outline the standard FMM approach using multipole expansions in Cartesian coordinates.
We employ tensor index notation with Latin indices ($i,j,k,...$) from the middle of the alphabet, and the unit metric $g_{ij}\equiv\delta_{ij}$.
For such metric, covariant and contravariant components are the same and there is no need to distinguish lower and upper indices.
The Latin indices from the beginning of the alphabet ($a$, $b$, $\dots$) will be used to enumerate particles. We also adopt the Einstein summation convention (repeated indices indicate a sum over that index, $\sum_i a^i b_i = a^i b_i$).

Computing gravitational forces for $N$ particles via pairwise interactions results in an algorithm that scales as $O(N^2)$ and quickly becomes computationally unfeasible for growing particle numbers.
FMM offers a solution to this problem by reducing this complexity to almost an $O(N)$ level~\cite{Cheng1999, Capuzzo-Dolcetta1998}.
This is achieved by replacing pairwise particle interactions with interactions between distant nodes in a particle tree, where a cluster of distant particles contained in a node is represented with a single point mass.
With that, pairwise particle interactions can be replaced with long-range interactions between nodes, provided that the latter are well-separated.
A gravitational potential of a remote node is approximated by a potential of a single point with gravitational monopole (mass), quadrupole, octopole, and higher moments up to desired precision.

\begin{figure}[htp]
  \begin{center}
    \includegraphics[width=0.94\columnwidth]{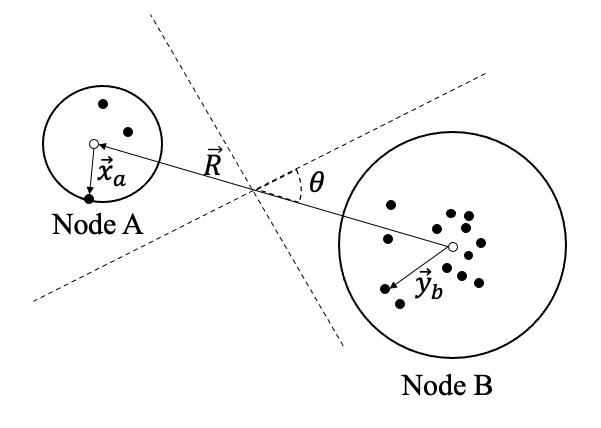}
  \end{center}
  \caption{Illustration of the FMM method. Depicted are two ``well-separated''
  nodes $A$ and $B$, each containing a cluster of particles. Gravitational field
  of all the particles inside $B$ is approximated by a field of a point mass $M_B$
  with optional quadrupole, octopole etc. The field at $\vec{x}_a$ is obtained by
  using the truncated Taylor expansion of the field from the center of mass of $A$.
  The angle $\theta$ shows the MAC angle.}
  \label{fig:main_fmm_sketch}
\end{figure}

The remarkable efficiency of FMM is complemented by the property of exact momentum conservation.
A gravitational field of the remote node $B$ acting on a particle $a$ inside the node $A$ is computed by the Taylor expansion of the gravitational potential from the center of mass (COM) of $A$ to the position of $a$ (see Figure \ref{fig:main_fmm_sketch}).
For conservation to work, the truncation order of the Taylor expansion must match the order of multipole moments retained in FMM.

In the following, we follow the presentation by Dehnen\cite{dehnen00}.
Consider two nodes, $A$ and $B$, with their respective COMs at $\ZA$ and $\ZB$, and let $\Rb := \ZA - \ZB$.
Let $\Xa$ and $\Yb$ be the locations of particles $a$ and $b$ within the nodes $A$ and $B$, with masses $m_a$ and $m_b$, respectively.
The gravitational potential $\ji_{a\leftarrow B}$ at $\Xa$ created by the node $B$ is due to the sum of individual interactions with other particles in $B$:
\begin{align}
    \ji_{a\leftarrow B} = -\sum_{b\in B} \GN\frac{m_b}{|\Xa - \Yb|},
    \label{eq:jia}
\end{align}
where $\GN$ is the Newtonian gravitational constant.
This is an exact expression, however, it scales as $O(N^2)$.

In FMM, we rewrite the expression $\Xa - \Yb = \Rb + (\xa - \yb)$ with $\xa := \Xa-\ZA$ and $\yb := \Yb-\ZB$ (see Figure~\ref{fig:main_fmm_sketch}) and expand the potential at $\Xa$ in Taylor series around the COM of node $A$:
\begin{align}
  \ji_{a\leftarrow b} &= \ji(|\Xa -\Yb|) =
  \nonumber\\
  &= \ji(R) + (\xa-\yb)\cdot \nabla\ji(|\bf{r}|)_{\bf{r}=\Rb} + \dots =
  \nonumber\\
  &= \sum_p \frac{1}{p!}
  \left[ [(\xa-\yb) \cdot \nabla ]^p
  \ji(|\bf{r}|)\right]_{\bf{r}=\Rb}.
\label{eq:taylor}
\end{align}

The series converges if $|\xa-\yb|$ is small: ${|\xa-\yb|<R}$.
This condition is always satisfied if the nodes are ``well-separated'', namely when their size is smaller than the distance between their COMs:
\begin{align}
    \frac{r_{A,{\rm max}} + r_{B,{\rm max}}}{R} \leq \tan\theta,
    \label{eq:mac_criterion}
\end{align}
where $r_{A,\textrm{max}}$ and $r_{B,\textrm{max}}$ are the maximum distances
from particles within the nodes to their respective COMs, and $\theta$ is the MAC angle. The latter controls whether the tree nodes are suitable to use multipoles. The series converges if $0 \leq \tan\theta < 1$ (see Fig.~\ref{fig:main_fmm_sketch}).

We can truncate the series at a finite $p$ to obtain an approximation for the potential to arbitrary precision.
The symmetry of the potential with respect to flipping the particles ${a\leftrightarrow b}$ guarantees that the forces will be antisymmetric (equal in magnitude and opposite in direction), thus producing exact conservation of linear momentum.

When summed over all particles in node $B$, the potential can be expressed in terms of the multipoles of $B$:
\begin{align}
   \ji_{a\leftarrow B} &= \sum_{b\in B}\ji_{a\leftarrow b} =
   \nonumber \\
   &= \sum_{b\in B}\ji(R) + \sum_{b\in B}(\xa-\yb)\cdot \nabla\ji(|\bf{r}|)_{\bf{r}=\Rb} + \dots =
   \nonumber \\
   &= -\frac{\GN M_B}{R} - \GN\Bigg\{
        \frac12 Q^{(B)}_{ij}  \frac{r^i r^j}{R^5}
        + \frac16 H^{(B)}_{ijk} \frac{r^i r^j r^k}{R^7} + \nonumber \\
      & \qquad\qquad\qquad + \frac1{24} X^{(B)}_{ijkl} \frac{r^i r^j r^k r^l}{R^9}
      + \dots
    \Bigg\},
   \label{eq:jiaB}
\end{align}
where $M_B$ is the node mass, $Q^{(B)}_{ij}$, $H^{(B)}_{ijk}$, and $X^{(B)}_{ijkl}$ are its quadrupole, octopole, and hexadecapole moments, respectively.
Here, the components $\{r^i\}$ represent vector $\xa$, with ${r\equiv|\xa|}$ being its magnitude.

So, a field created at the location of a particle $a$ by a node $B$ is given by a relatively simple and fast-to-evaluate expression (\ref{eq:jiaB}), representing a truncated Taylor series.
We can, therefore, compute gravitational forces between the nodes $A$ and $B$ in two steps: (i) compute multipole moments of the nodes; (ii) apply Taylor expansion from their respective COM to the particle locations.
Summing the symmetric expression (\ref{eq:taylor}) over the particles in the node $B$ gives the Taylor expansion from COM of node $A$, and vice versa.
As long as both Taylor expansions are obtained from the same symmetric expression, the total gravitational forces cancel out exactly.

However, this method also introduces torques generated between pairs of particles, which in the case of exact $O(N^2)$ method vanish identically.
Indeed, even though the forces are equal in magnitude and opposite in direction, they are not necessarily aligned with the line connecting particles $a$ and $b$, producing small nonzero torques.
The existence of these torques leads to violation of the angular momentum conservation.
A possible solution was found by Marcello~\cite{marcello17} by introducing artificial compensating torques to counteract the spurious FMM torques between pairs of nodes.\\
Let us illustrate linear momentum conservation versus angular momentum nonconservation in FMM using the Taylor expansion up to 4th order.

\subsection{Example: expansion up to 4-th order}
\label{sec:sym_fmm}

The monopole $M$ and the quadrupole moments $Q_{ij}$ are given by
\begin{align}
\label{eqn:quad_mom}
M &= \sum_b m_b, \\
Q^{ij} &= \sum_b m_b (3 y_b^i y_b^j - \delta_{ij} q_b^2),
\end{align}
where $y_b^i = Y^i_b - Z_B^i$ (or, in vector notation,  ${\bf y}_b := {\bf Y}_b - \ZB$), and $q_b := |{\bf y}_b|$.

The octopole moment $H_{ijk}$ is given by
\begin{align}
H^{ijk} = \sum_b m_b \left[15 y_b^i y_b^j y_b^k - 3 q_b^2(\delta_{ij} y_b^k + \delta_{jk} y_b^i + \delta_{ik} y_b^j) \right] .
\label{eqn:octo_mom}
\end{align}
The hexadecapole moment $X_{ijkl}$ is given by
\begin{align}
\label{eqn:hexa_mom}
X^{ijkl} &= \sum_b m_b \big[105 y_b^i y_b^j y_b^k y_b^l
	     - 15 q_b^2(\delta_{ij} y_b^k y_b^l +  \delta_{il} y_b^j y_b^k \nonumber \\
	     &\qquad \qquad +\delta_{ik} y_b^j y_b^l
	                 + \delta_{jl} y_b^i y_b^k + \delta_{jk} y_b^i y_b^l +\delta_{lk} y_b^i y_b^j)\nonumber \\
	      &\qquad \qquad +3q_b^4(\delta_{ij}\delta_{kl}+\delta_{ik}\delta_{jl}+\delta_{il}\delta_{jk})\big] .
\end{align}

The gradient and higher partial derivatives of the gravitational potential $\ji$, $\nabla^p \ji(R) = -\nabla^p \frac{1}{R}$, are \cite{marcello17}:
\begin{align}
    D &= -\frac{1}{R} \; , \label{eqn:D} \\
    D_i &= \frac{R^i}{R^3} \; ,\label{eqn:Di}\\
    D_{ij} &= \frac{-3 R^i R^j + \delta_{ij} R^2}{R^5} \; , \label{eqn:Dij}\\
    D_{ijk} &= \frac{15 R^i R^j R^k - 3 R^2(\delta_{ij} R^k + \delta_{jk} R^i + \delta_{ik} R^j)}{R^7} \; ,\label{eRn:Dijk}
\end{align}
\begin{align}
    D_{ijkl} &= \frac{1}{R^9} \big[-105 R^i R^j R^k R^l + \nonumber \\
	     &\qquad + 15 R^2(\delta_{ij} R^k R^l + \delta_{il} R^j R^k
	            + \delta_{ik} R^j R^l +\nonumber \\
	     &\qquad\qquad+ \delta_{jl} R^i R^k
	            + \delta_{jk} R^i R^l + \delta_{lk} R^i R^j) \nonumber \\
	      &\qquad -3R^4(\delta_{ij}\delta_{kl}+\delta_{ik}\delta_{jl}+\delta_{il}\delta_{jk})\big] . \label{eqn:Dijk}
\end{align}
The 4-th order truncated Taylor expansion of the gravitational potential at $\Xa$ generated by the node $B$ gives:
\begin{align}
&\ji_{a\leftarrow B}(\Xa) \approx M_B D + D_i M_B x^i
    + \frac12 D_{ij}(Q^{ij}_B + M_B x^i x^j)  \nonumber \\
   &+ \frac16 D_{ijk}(H^{ijk}_B + 3 Q^{ij}_B x^k + M_B x^i x^j x^k)  \nonumber \\
   &+ \frac{1}{24} D_{ijkl} (X^{ijkl}_B + 4 H^{ijk}_B x^l + 6 Q^{ij}_B x^k x^l + M_B x^i x^j x^k x^l) .
    \label{eqn:potential}
\end{align}
Using Eqn.~\ref{eqn:potential}, the gravitational acceleration (specific force) exerted by the node $B$ at a point $\Xa$ in the node $A$ can be computed as follows (using $\vec{a} = -\nabla\ji$):
\begin{align}
a^i_{a\leftarrow B}(\Xa) &\approx -M_B D_i - D_{ij} M_B x^j  \nonumber \\
    &-\frac12 D_{ijk}(Q^{jk}_B + M_B x^j x^k)  \nonumber \\
    &-\frac16 D_{ijkl}(H^{ijk}_B + 3 Q^{jk}_B x^l + M_B x^j x^k x^l).
    \label{eqn:g_acc_BtoA}
\end{align}
It is straightforward to check that the net force between the nodes $A$ and $B$ vanishes identically:
\begin{align}
    \label{eqn:netforce}
    \Fb_{B \leftarrow A} + \Fb_{A \leftarrow B}
    = \sum_{a\in A} m_a {\bf a}_{a\leftarrow B}
    + \sum_{b\in B} m_b {\bf a}_{b\leftarrow A}
    = 0.
\end{align}
Zero net force implies exact conservation for linear momentum.
For the angular momentum, on the other hand, this result does not hold.
The net torque $\mathbf{\tau}_{AB}$ taken at the COM of the node $B$ due to the mutual gravitation of $A$ and $B$ is:
\begin{align}
    \label{eqn:total_torque}
    \tau_{AB}^i =
         \sum_{a\in A} m_a \epsilon_{ijk} (R^j + x_a^j) a^k_{a\leftarrow B}
       + \sum_{b\in B} m_b \epsilon_{ijk} y_b^j a^k_{b\leftarrow A} ,
\end{align}
where $\epsilon_{ijk}$ is the Levi-Civita symbol. Using the previous expressions for acceleration, we can obtain the net nonzero torque up to 4-th order (check also  \cite{marcello17} for a simplified result):
\begin{align}
    \label{eqn:total_torque4th}
    \tau^i_{AB} &\approx \epsilon_{ijk}D_{kl}(M_A Q_B^{jl} - M_B Q_A^{jl} ) \nonumber \\
    &+\frac{1}{2}\epsilon_{ijk}D_{klm}(M_A H_B^{jlm} - M_B H_A^{jlm} ) \nonumber \\
    &+\frac{1}{6} \epsilon_{ijk}D_{klmp}\Big[3 ( Q_A^{mp} Q_B^{jl} -  Q_B^{mp} Q_A^{jl} ) \nonumber\\
    &\quad + M_A X_B^{jlmp} - M_B X_A^{jlmp} \Big] \nonumber \\
    &-\frac{1}{2} \epsilon_{ijk} D_{klm} (M_A Q_B^{lm} + M_B Q_A^{lm}) R^j \nonumber \\
    &-\frac{1}{6} \epsilon_{ijk} D_{klmp} (M_A H_B^{lmp} + M_B H_A^{lmp}) R^j .
\end{align}

\section{Improving the Method}
\label{sec:solution}
We consider two different approaches to conserve angular momentum in FMM.

\subsection{Zeroth Order FMM}
\label{sec:zerothFMM}

One of the simplest solutions to conserve angular momentum in FMM is to refrain from using FMM orders higher than zeroth.
Indeed, when only the constant term is left in the expression for acceleration (\ref{eqn:g_acc_BtoA}), there is no net torque.
Every particle experiences the same acceleration from the other node, both in magnitude and in direction. As a result, the net spin torque on both nodes is zero.
The torque between COMs of the nodes is also zero, because the action and reaction are exactly aligned along the line connecting the COMs.
This can also be inferred from the expression (\ref{eqn:total_torque4th}) where there is no monopole contribution in the torques.

The problem with low-order expansion is that it lacks accuracy.
The accuracy can be improved by decreasing MAC, but this requires compromising the efficiency.
On the other hand, simplicity of implementation makes this approach ideal for a class of problems where angular momentum conservation may be more important than the accuracy of interactions.

\subsection{Realigning pairwise forces}

In the exact Newtonian gravity of $N$ interacting particles, conservation of angular momentum comes not only from the balance of action and reaction in pairwise interactions but also from the fact that they are collinear with the line connecting the particles.
For the same reason, in FMM non-conservation of angular momentum ultimately comes from the misalignment of the pairwise forces between particles on different nodes, as approximated by the truncated Taylor series (\ref{eq:taylor}).
One way to restore it, therefore, would be to re-align these forces, for instance by projecting them onto the unit vector $\nb_{ab}$ in the direction from $a$ to $b$:
\begin{align}
    \ab^{\|}_{a\leftarrow b} = (\ab_{a\leftarrow b} \cdot \nb_{ab}) \nb_{ab}.
    \label{eq:realign}
\end{align}
Another possibility is to simply multiply the unit vector $\nb_{ab}$ by the acceleration magnitude:
\begin{align}
    \ab^{(0),\|}_{a\leftarrow b} = -|\ab^{(0)}_{a\leftarrow b}| \nb_{ab}.
    \label{eq:realign2}
\end{align}
The unit vector $\nb_{ab}$ in (\ref{eq:realign}) and (\ref{eq:realign2}) can be expressed in terms of ${\nb_R := \Rb/R}$, the unit vector along the line connecting COMs of the nodes:
\begin{align}
    \nb_{ab} = \frac{\Xa - \Yb}{|\Xa - \Yb|}
    = \frac{\nb_R + \db_{ab}}{|\nb_R + \db_{ab}|},
    \label{eqn:nab1}
\end{align}
where $\db_{ab} := (\xa - \yb)/R$ must be small such that $\nb_{ab}$ can be expanded around $\nb_R$ in powers of $\db_{ab}\propto\xa - \yb$, in the spirit of FMM:
\begin{align}
    &\nb_{ab} \approx (\nb_R + \db_{ab})(1 - \nb_R\cdot\db_{ab} + \dots) =
    \nonumber \\
    &\qquad= \nb_R + \db_{ab} - (\nb_R\cdot\db_{ab})\nb_R  - (\nb_R\cdot\db_{ab})\db_{ab},
    \label{eq:nab-expansion} \\
    &(\ab_{a\leftarrow b}\cdot\nb_{ab}) \nb_{ab}
    = \frac{(\ab_{a\leftarrow b},\nb_R + \db_{ab})}{|\nb_R + \db_{ab}|^2}
    (\nb_R + \db_{ab}) \approx
    \nonumber \\
    &\approx
    \bigg[(\ab_{a\leftarrow b}\cdot\nb_R)
    + (\ab_{a\leftarrow b}\cdot\db_{ab})
    \nonumber \\
    &\qquad - 2(\ab_{a\leftarrow b}\cdot\nb_R)(\db_{ab}\cdot\nb_R) + \dots
    \bigg]
    (\nb_R + \db_{ab}).
    \label{eq:abnab-expansion}
\end{align}
Both expressions (\ref{eq:nab-expansion}) and (\ref{eq:abnab-expansion}) have the form of a scalar quantity times $(\nb_R+\db_{ab})$, which is collinear with $\nb_{ab}$.
In these scalar quantities, we only keep terms up to the first order, but higher terms can be added for better accuracy.
The overall expressions (\ref{eq:nab-expansion}) and (\ref{eq:abnab-expansion}) are quadratic in $\db_{ab}$, so, when summed over all particles $b\in B$, they produce terms up to a quadrupole.

At zeroth order, the net acceleration on a particle $a$ from the node $B$ according to equations (\ref{eq:realign2}) and (\ref{eq:nab-expansion}) is: \begin{align}
\ab^{(0),\|}_{a\leftarrow B}
&= \sum_{b\in B} \ab^{(0),\|}_{a\leftarrow b}
= -|\nabla\ji^{(0)}(R)| \sum_{b\in B}m_b\nb_{ab}
\nonumber \\
&\approx -\frac{G}{R^2}\sum_{b\in B} m_b
\bigg[\nb_R + \db_{ab} -
\nonumber \\
 &- (\nb_R\cdot\db_{ab})\nb_R
  - (\nb_R\cdot\db_{ab})\db_{ab}\bigg] =
\nonumber\\
&= -\frac{GM_B}{R^2}\bigg[ \nb_R
+ \frac{1}{R}\xa^\perp -
\nonumber \\
&\quad-\frac{1}{R^2}(\xa\cdot\nb_R)\xa
-\frac{1}{M_B R^2}\nb_R\cdot\bar{\bar{Q}}_B\bigg],
\label{eq:aaB0}
\end{align}
where $\xa^\perp := \xa - \nb_R(\nb_R\cdot\xa)$ is the component of $\xa$ orthogonal to $\nb_R$.
Since we made an approximation (\ref{eq:nab-expansion}), the length of the vector on which we are projecting is incorrect. However, because its direction is along the line connecting the two particles, the artificial torque automatically vanishes, and angular momentum conservation is achieved.
Moreover, because the projected pairwise force is antisymmetric in the particles $a$ and $b$, the linear momentum conservation is also respected.

The first three terms inside the square brackets in (\ref{eq:aaB0}) represent a force reoriented towards the COM of $B$, and the last term is a quadrupole interaction:
\begin{align}
    \ab^{(0),\|}_{a\leftarrow B}
    \approx -\frac{GM_B}{R^2}\nb_{aB}
    + \frac{G}{R^4}\bar{\bar{Q}}_B\cdot\nb_R.
\label{eq:aab0-approx}
\end{align}

\begin{figure*}[htp!]
\begin{tabular}{cc}
  \includegraphics[width=0.47\textwidth]{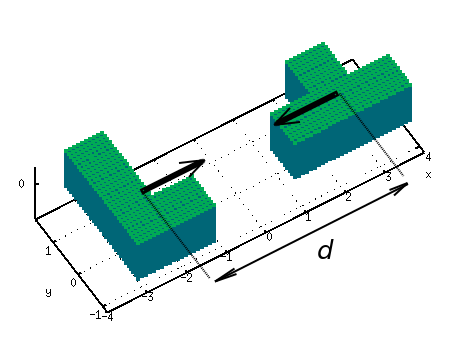} &
  \includegraphics[width=0.47\textwidth]{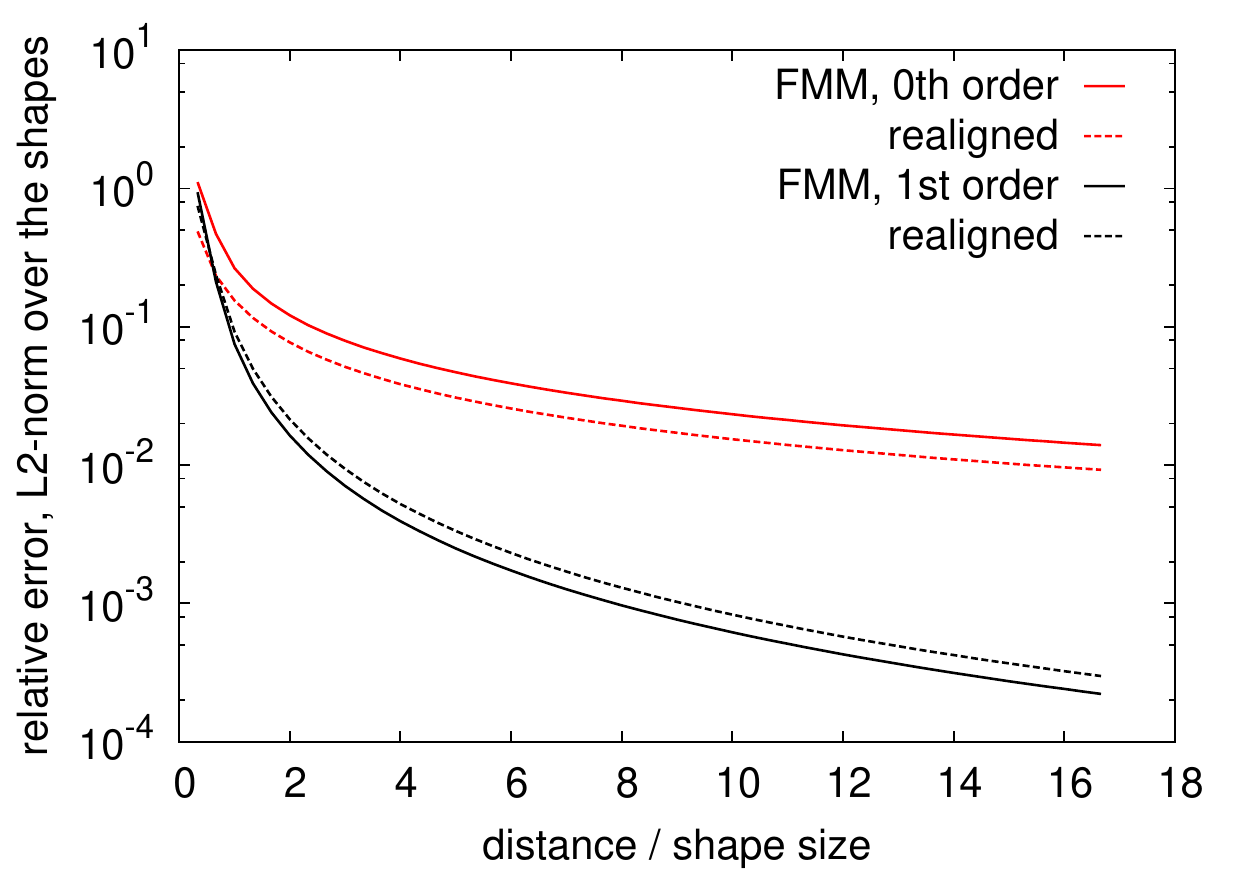}
\end{tabular}
\caption{Left panel: setup for a simple test. Two shapes are placed at a varying
distance $d$ between their centers of mass. The shapes consist of $4\times10^3$
particles each.
Right panel: relative error between the exact Newtonian force and approximate force,
as a function of distance, computed using the 0th and 1st order FMM, with and
without realignment. The error is measured for every particle and averaged over the
shapes using the $L_2$ norm.}
\label{fig:realign}
\end{figure*}

However, as discussed above, zeroth-order FMM already conserves angular momentum. Non-conservation issues appear at the first order and higher.
With the notation $\xib:=\xa/R$, $\etab:=\yb/R$, and the matrix $\bar{\bar{D}}:=\|D_{ij}\|$ from Eq.(\ref{eqn:Dij}), the first-order correction for the acceleration of a particle $a$ due to particle $b$ is:
\begin{align}
    \ab^{(1)}_{a\leftarrow b} = -m_b R \bar{\bar{D}}\cdot\xib.
\end{align}
In the subsequent derivation, it is convenient to introduce a ``scalar product'' based on the matrix $\bar{\bar{D}}$:
\begin{align}
    \langle\boldsymbol{\xi},\boldsymbol{\eta}\rangle_D := R D_{ij}\xi^i\eta^j.
\end{align}
To get the net realigned first-order correction, we can use (\ref{eq:realign}) and (\ref{eq:abnab-expansion}):
\begin{align}
    &\ab^{(0),\|}_{a\leftarrow B} + \ab^{(1),\|}_{a\leftarrow B}
    = \sum_{b\in B} (\ab^{(0)}_{a\leftarrow b} + \ab^{(1)}_{a\leftarrow b}\cdot\nb_{ab})\nb_{ab} \approx
    \nonumber \\
    &\approx -M_b \big[D^2 (1 - \xib\cdot\nb_R) + \langle\xib,\nb_R\rangle_D \big](\nb_R + \xib)
    \nonumber \\
    &\quad- \langle\nb_R,\bar{\bar{Q}}\rangle_D D^2 + (\nb_R\cdot\bar{\bar Q}_B) D^4 .
\label{eq:aaB1}
\end{align}
The equation (\ref{eq:aaB1}) gives an expression for the net acceleration, accurate up to $O(\delta^2)$ and conserving both linear and angular momenta.

\section{Results}
\label{sec:results}

A simple test of the new method is presented in Figure~\ref{fig:realign}.
In the test, we explored the error in computing the Newtonian force acting on particles arranged in two irregular shapes.
Figure~\ref{fig:realign} shows the error as a function of distance between the shapes.
The distance is shown in the units of characteristic shape size (3~cm).
Red and black curves display errors for zeroth and first-order FMM, with (dashed) and without (solid) realignment.
It demonstrates that the accuracy of the FMM method is not affected. This is expected, since the expansions (\ref{eq:aaB0}) and (\ref{eq:aaB1}) in our realignment method are carried out to second order in $\delta$, $O(\delta^2)$, higher than the FMM orders to which it is applied.
Most importantly, both the net force and the net torque on the particles vanish at machine precision.

The zeroth order FMM method has been implemented in \texttt{FleCSPH}~\cite{Loiseau2020}. \texttt{FleCSPH}\footnote{https://github.com/laristra/flecsph} is a smoothed particle hydrodynamics simulation tool based on \texttt{FleCSI}~\cite{charest2017flexible}.
\texttt{FleCSI}\footnote{https://github.com/laristra/flecsi} is a compile-time configurable framework providing distributed and parallel topologies, such as structured and unstructured mesh, Narray, and Ntree.
\texttt{FleCSPH} is based on the MPI implementation of the Ntree topology implementing distributed binary, quad, and octree in 1, 2, and 3 dimensions respectively.

To demonstrate the implementation, we test evolution of a stable isolated star
in equilibrium. This test checks consistency and conservation properties for the coupled hydrodynamics and gravity.

\begin{figure}[h]
\includegraphics[width=0.45\textwidth]{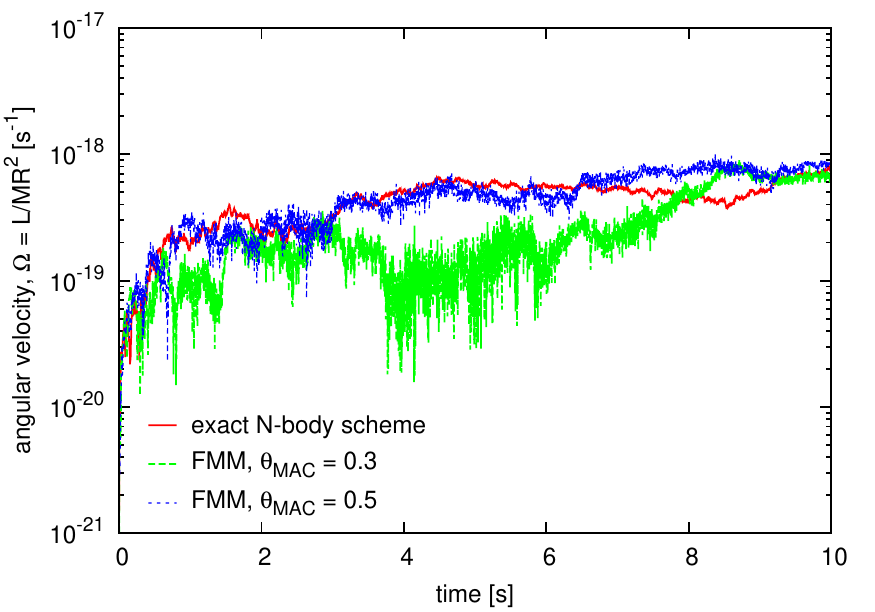}
\caption{Evolution of angular momentum for a self-gravitating
model of a white dwarf with two different MAC values: $\tan \theta_{\textrm{MAC}} = 0.3$, and $0.5$.}
\label{fig:WD_fmm_test}
\end{figure}

For the initial data, we solve Lane-Emden equation~\cite{chandra1957}
for polytrope with $\Gamma=5/3$, $K = 10^{12}$ (in CGS units), and central density $\rho_c =5.2\times10^{6}\,{\rm g}\,{\rm cm}^{-3}$.
This results in a polytrope resembling a white dwarf, with mass $0.2\,M_\odot$ and radius 4790~km.

Fig.~\ref{fig:WD_fmm_test} shows the evolution of angular momentum for a self-gravitating model of a white dwarf, with zeroth-order FMM method.
Two different MAC angles, $\tan \theta_{\textrm{MAC}} = 0.3$ and $0.5$, were chosen and compared
with exact $N$-body scheme. Both methods conserve angular momentum up to machine precision, as expected.

\section{Conclusion and Discussion}
\label{sec:conclusion}
We present a new modification to the FMM that allows to conserve angular momentum by construction.
The main idea of the method is to realign the approximate pairwise forces in such a way that they are parallel to the line connecting two interacting particles.
The unit vector along this direction can be expanded in terms of the difference between the local particle position vectors relative to the COMs of their nodes, ${\db\propto(\xa-\yb)}$, and then resummed to arbitrary multipole order, while retaining the momentum conservation property in the spirit of the standard FMM.
The method is demonstrated for zeroth and first orders with a simple setup of gravitational interaction between two irregular shapes.
We also show an excellent conservation of angular momentum in the evolution of a single star with zeroth order FMM (see Figure~\ref{fig:WD_fmm_test}).

The method can be extended to higher multipole moments. The underlying algebra of multipole operators is that of symmetric tensors, which can be compactly represented. Software that manipulates and translates symbolic expressions, such as SymPy \cite{meurer2017sympy}, can greatly simplify the generation of code to evaluate multipole operators. We will therefore seek opportunities to automate our method before advancing to higher multipoles case in the future work.

\appendices

\section*{Acknowledgment}
The authors would like to thank Christopher Fryer and Stephan Rosswog for valuable feedback that helped improve the paper. This work is supported by the LANL ASC Program and LDRD grants 20200145ER and 20190021DR. This work used resources provided by the LANL Institutional Computing Program. LANL is operated by Triad National Security, LLC, for the National Nuclear Security Administration of the U.S.DOE  (Contract No. 89233218CNA000001). This work is authorized for unlimited release under LA-UR-21-24198.



\end{document}